\documentclass{PoS}
\usepackage{graphicx} 
\usepackage{floatrow}
\usepackage{lineno}

\title{Towards a W boson mass measurement with LHCb}

\ShortTitle{Towards a \boldmath{$W$} boson mass measurement with LHCb}

\author{\speaker{Martina Pili}\thanks{On behalf of the LHCb Collaboration.}\\
        University of Oxford\\
        E-mail: \email{martina.pili@cern.ch}}

\abstract{LHCb provides unique opportunities to study $W$ and $Z$ boson production at forward rapidities at the LHC.
It has recently been suggested that a measurement of the $W$ boson mass by LHCb would complement measurements by ATLAS and CMS.
All measurements of the $W$ mass at hadron colliders are subject to PDF uncertainties, but there would be a partial cancellation of the overall PDF uncertainty when the LHCb result is included in an average with measurements by ATLAS and CMS.
Here we review measurements of $W$ and $Z$ boson production by LHCb, and report on a new study of the PDF uncertainty on the LHCb measurement of the $W$ mass.
The latter study includes the proposal of a new approach that should reduce the PDF uncertainty by roughly a factor of two with LHCb Run 2 data.}

\FullConference{XXVII International Workshop on Deep-Inelastic Scattering and Related Subjects - DIS2019\\
		8-12 April, 2019\\
		Torino, Italy}
		
\begin{document}

\section{Introduction}
Global fits to electroweak (EW) observables are a powerful probe of physics beyond the Standard Model (SM) but they are currently limited by the precision with which certain observables, like the $W$ boson mass ($M_W$), are measured.  
The first measurement of $M_W$ at the LHC by the ATLAS collaboration~\cite{ATLAS} is already competitive with results from the Tevatron~\cite{CDF,D0}, but the theoretical uncertainties in the $W$ production model, in particular those related to the PDFs, represent a limiting factor.
LHCb~\cite{LHCb} is a forward spectrometer with full charged particle tracking and identification capabilities over the range $2<\eta< 5$, almost orthogonal to those of ATLAS and CMS, which allows it to probe complementary phase space regions. Despite being primarily designed for the study of beauty and charm hadrons, LHCb has a strong track record in measurement of $W$ and $Z$ boson production in muonic final states~\cite{LHCbW,LHCbZ}. As far as precision EW tests are concerned, LHCb has already measured the value of the effective weak mixing angle~\cite{sin2theta} but the potential for a measurement of $M_W$ by LHCb was realised only recently. 
Ref.~\cite{MWprop} proposed a measurement of $M_W$ based on a template fit of the transverse momentum distribution of forward muons from $W$ decays. 
It is estimated that the Run 2 data could yield a $M_W$ measurement with a statistical uncertainty of roughly 10\,MeV/c$^2$. Ref.~\cite{MWprop} estimated that the PDF uncertainties in a standalone LHCb measurement would be larger than those in ATLAS and CMS. However, the uncertainty on the LHCb measurement would be partially anticorrelated with those of ATLAS and CMS. It is therefore claimed that the introduction of a LHCb measurement into a LHC $M_W$ average could reduce the overall PDF uncertainty.
Ref.~\cite{PDFpaper} aims to understand what drives the PDF uncertainty in LHCb and how it can be reduced.

\section{Fitting Method}
\begin{figure}
\floatbox[{\capbeside\thisfloatsetup{capbesideposition={right,center},capbesidewidth=7cm}}]{figure}[\FBwidth]
{\caption{Upper: the distribution of the best-fit $\chi^2$ versus $M_W$ for a fit to a single toy dataset from the $W^-$ sample, which assumes the LHCb Run 2 statistics, with each of the 1000 NNPDF3.1 replicas. Lower: the distribution of the $M_W$ values with a Gaussian fit function overlaid.}\label{fig:1}}
{\includegraphics[width=0.46\textwidth]{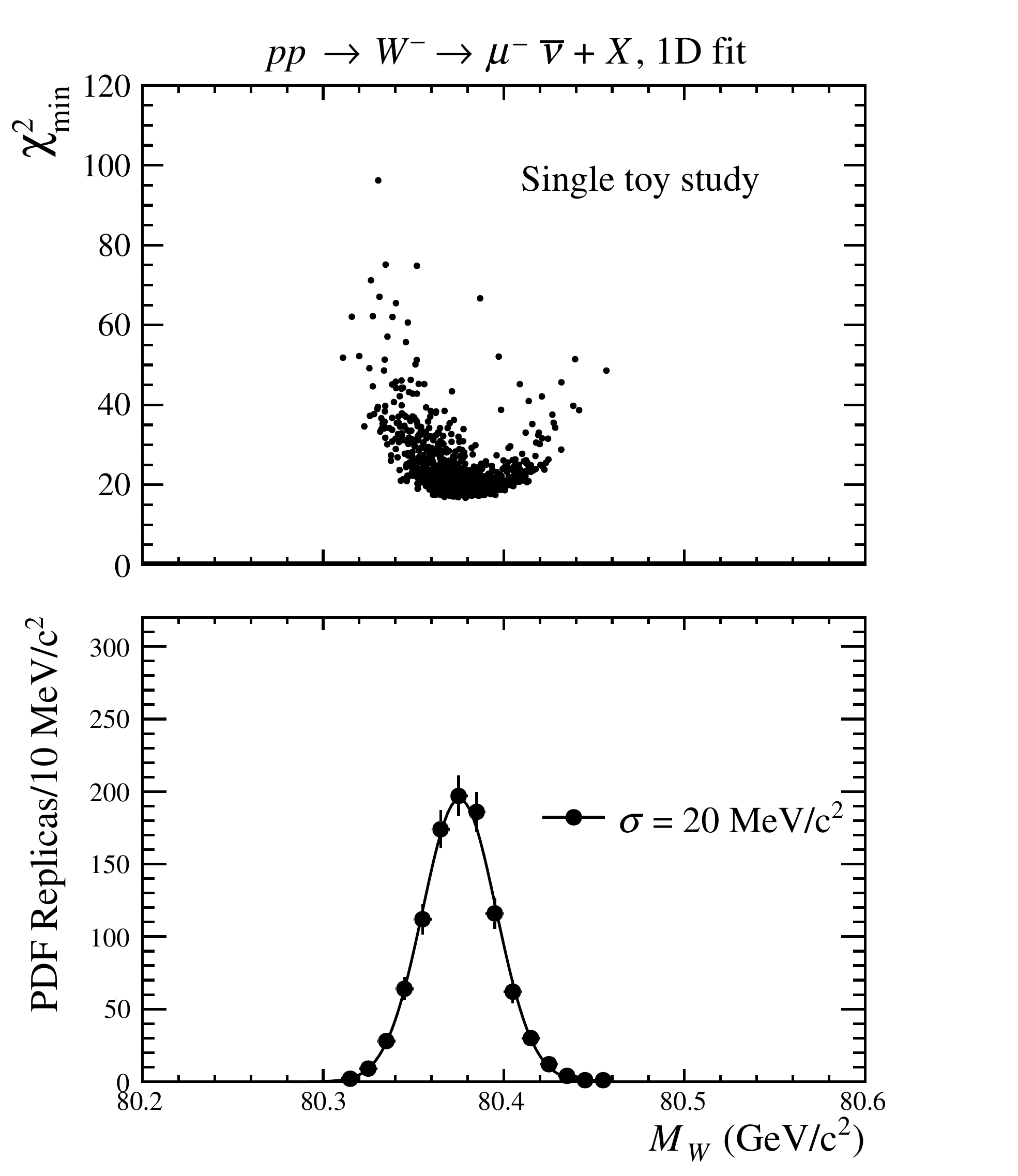}}
\end{figure}

Around $10^8$ events are generated using POWHEG~\cite{powheg} with a parton shower provided by Pythia~\cite{pythia} at center-of-mass energy $\sqrt{s}$ = 13\,TeV, and about 10 million events in the range 30 $<p_T^\mu < $ 50\,GeV/c and 2 $<\eta<$ 4.5 are selected. 
Toy data histograms are generated by randomly fluctuating the bins around the nominal distribution, assuming the expected Run 2 yields and Poisson statistics. The data histograms are compared to templates with different PDF and $M_W$ hypotheses. 
Ref.~\cite{PDFpaper} makes use of the NNPDF3.1~\cite{NNPDF} set with 1000 equiprobable replicas. 
For a given PDF hypothesis a single-parameter fit determines the value of $M_W$ that minimises the $\chi^2$ between the toy data and the templates. 
Fig.~\ref{fig:1} shows that the $M_W$ values extracted for multiple PDF hypotheses are approximately distributed according to a Gaussian and the PDF uncertainty is defined as its width. The broadly parabolic distribution of the best-fit $\chi^2$ ($\chi^2_{min}$) versus $M_W$ indicates that the PDF replicas that most severely bias $M_W$ tend to give a measurably poorer fit quality. This information can be used to constrain the PDF uncertainty but first, let us try to understand what drives it at LHCb. For visual purposes some sections report only results from the $W^-$ dataset, but the results for the $W^+$ are included in Ref.~\cite{PDFpaper}.

\section{Understanding the PDF uncertainties}
\begin{figure}
\centerline{\includegraphics[width=0.46\textwidth]{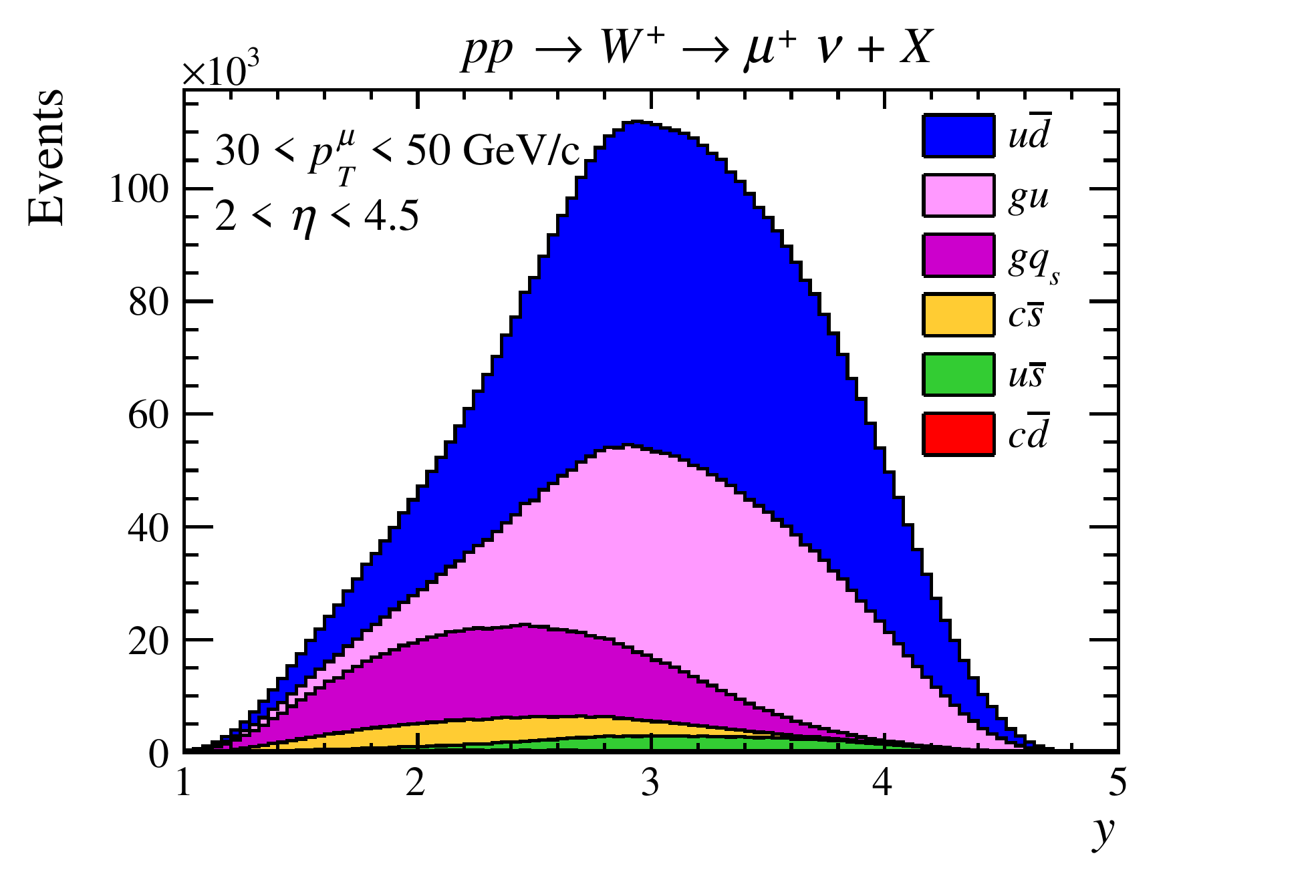}\includegraphics[width=0.46\textwidth]{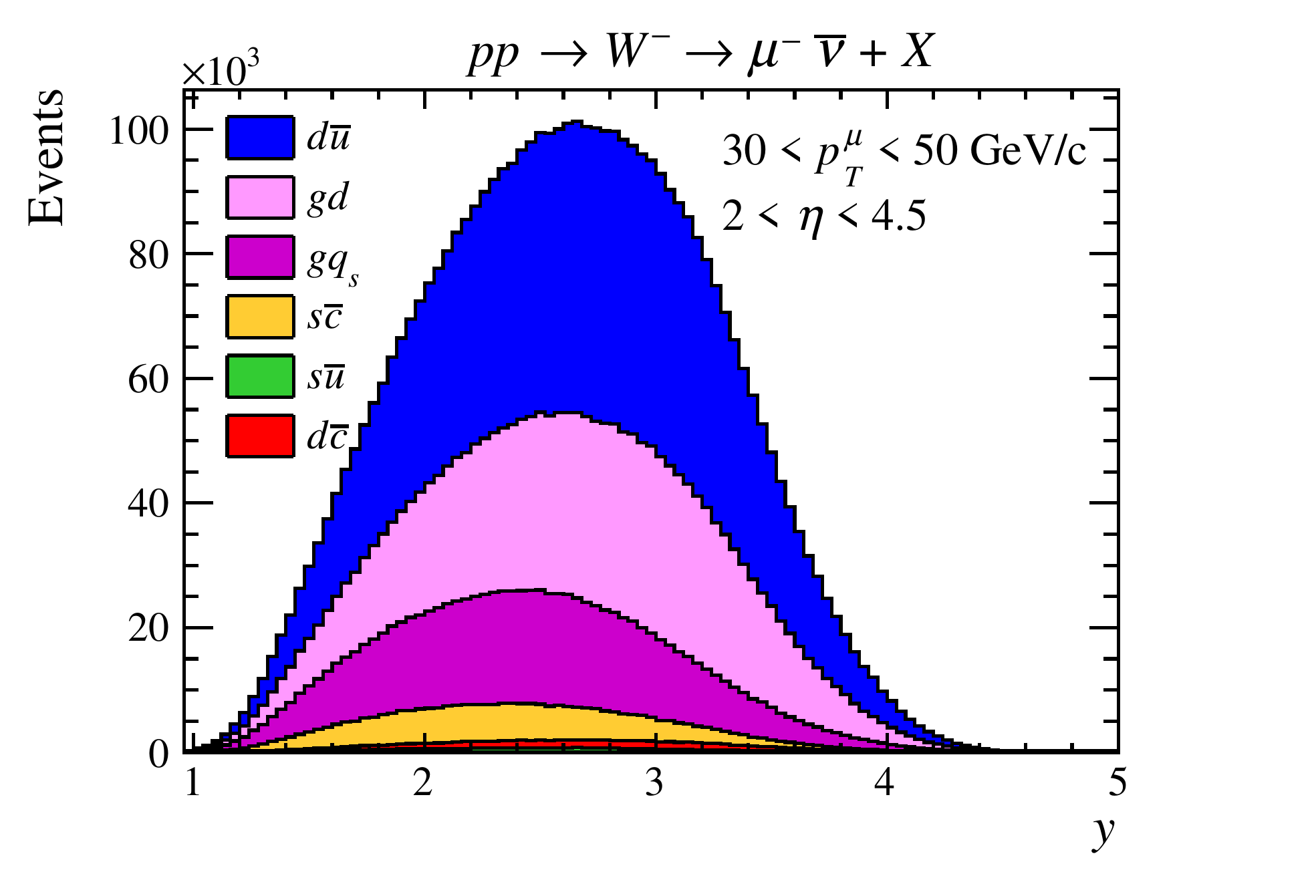}}
\caption{The (left) $W^+$ and (right) $W^+$ rapidity distributions decomposed into the main partonic subprocesses.}\label{fig:2}
\end{figure}

\begin{figure}
\centerline{
\includegraphics[width=0.46\textwidth]{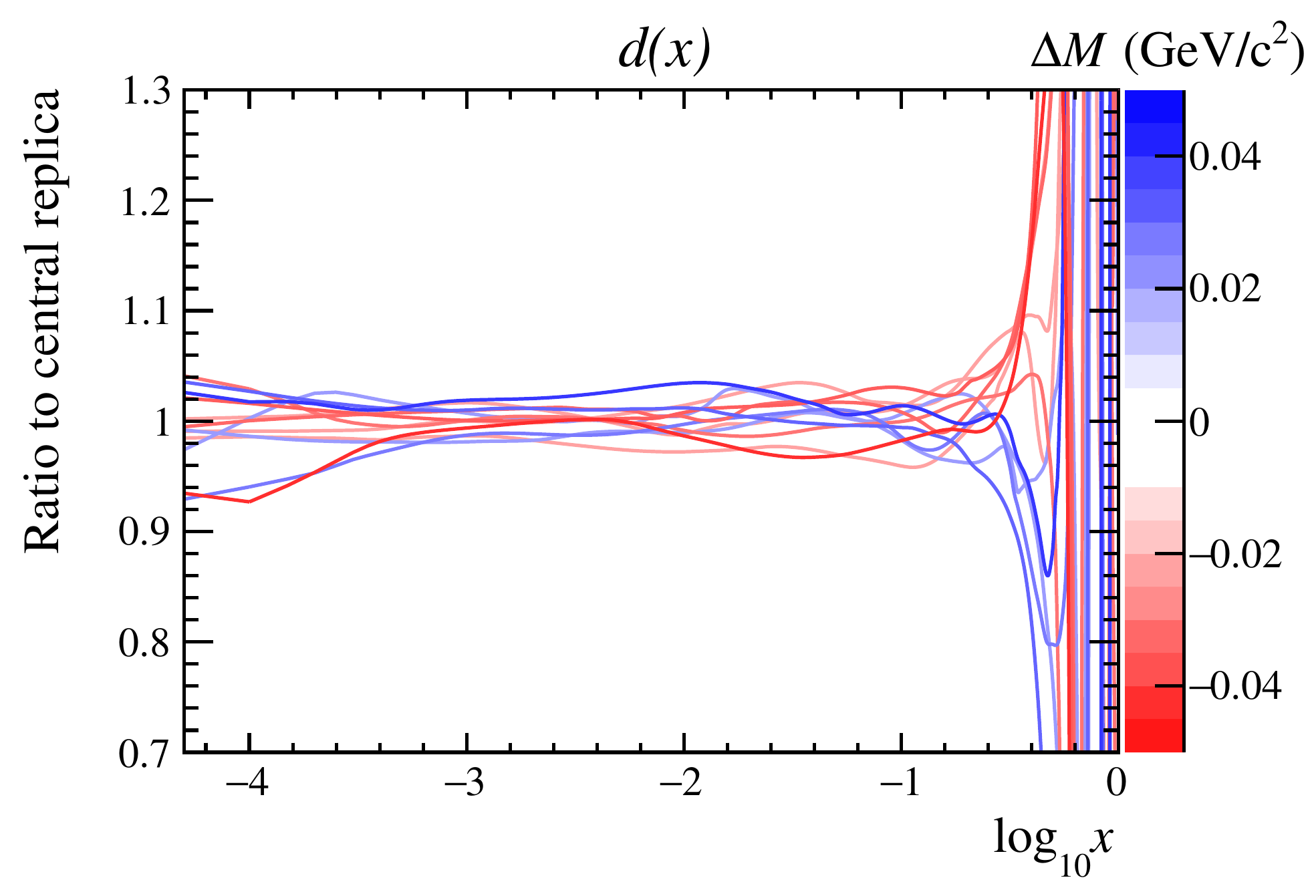}\includegraphics[width=0.46\textwidth]{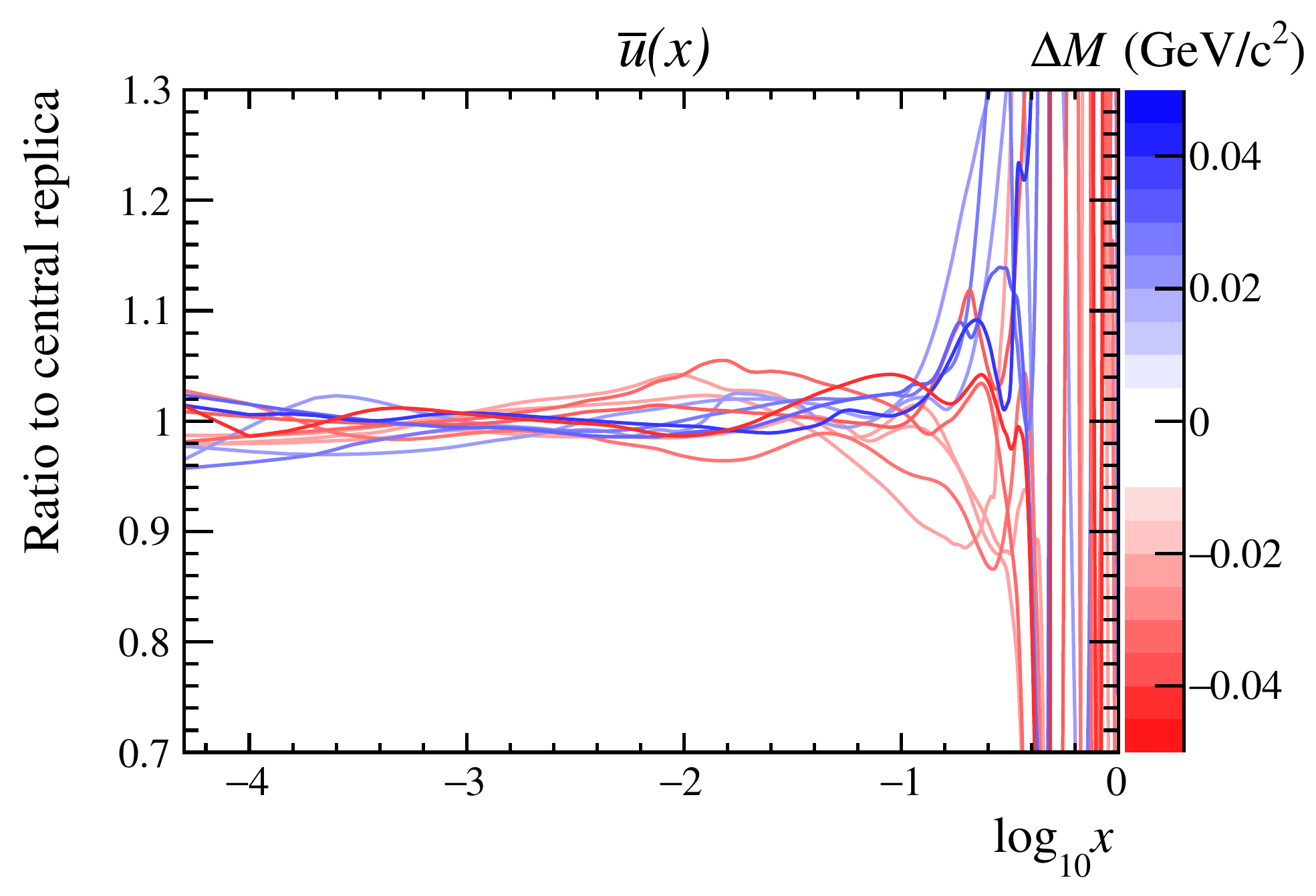}}
\caption{The ratios of a subset of NNPDF3.1 replicas with respect to the central replica, for the $x$ dependence of the $d$ and $\bar{u}$ PDFs. Each line is marked with a colour indicating the shift of the $M_W$ value determined from a fit to the $p_T^\mu$ distribution of a single toy dataset.}\label{fig:3}
\end{figure}

Fig.~\ref{fig:2} shows how the different partonic subprocesses contribute to the cross-section for $W$ production as a function of rapidity ($y$). The dominant $W^+$($W^-$) production subprocesses involve valence $u$($d$) quarks, with a roughly 20\% contribution from annihilation of gluons with sea quarks ($gq_s$). Contributions from the annihilation of second generation quarks are below 10\%.
Since the first quark generation seems to be the most important it is interesting to see if
there are any obvious patterns in the corresponding PDFs for the replicas that lead to biased $M_W$ determinations. The studies in this section make use of a subset of the NNPDF3.1 replicas. Fig.~\ref{fig:3} show how the $x$ dependencies of the $d$ and $\bar{u}$ PDFs vary between the subset of replicas. Each line is a ratio with respect to the central replica, and is assigned a colour according to the bias in $M_W$ as evaluated using the method described earlier. A clear pattern can be seen for the high-$x$ (above $x \approx 0.1$) $d$ PDF, whereby the replicas that tend to bias $M_W$ upwards (downwards) tend to have a smaller (larger) parton density. An opposite sign pattern is seen in the $\bar{u}$ PDF. No obvious pattern is observed in the $W^+$ case.

\begin{figure}
\centerline{\includegraphics[width=0.46\textwidth]{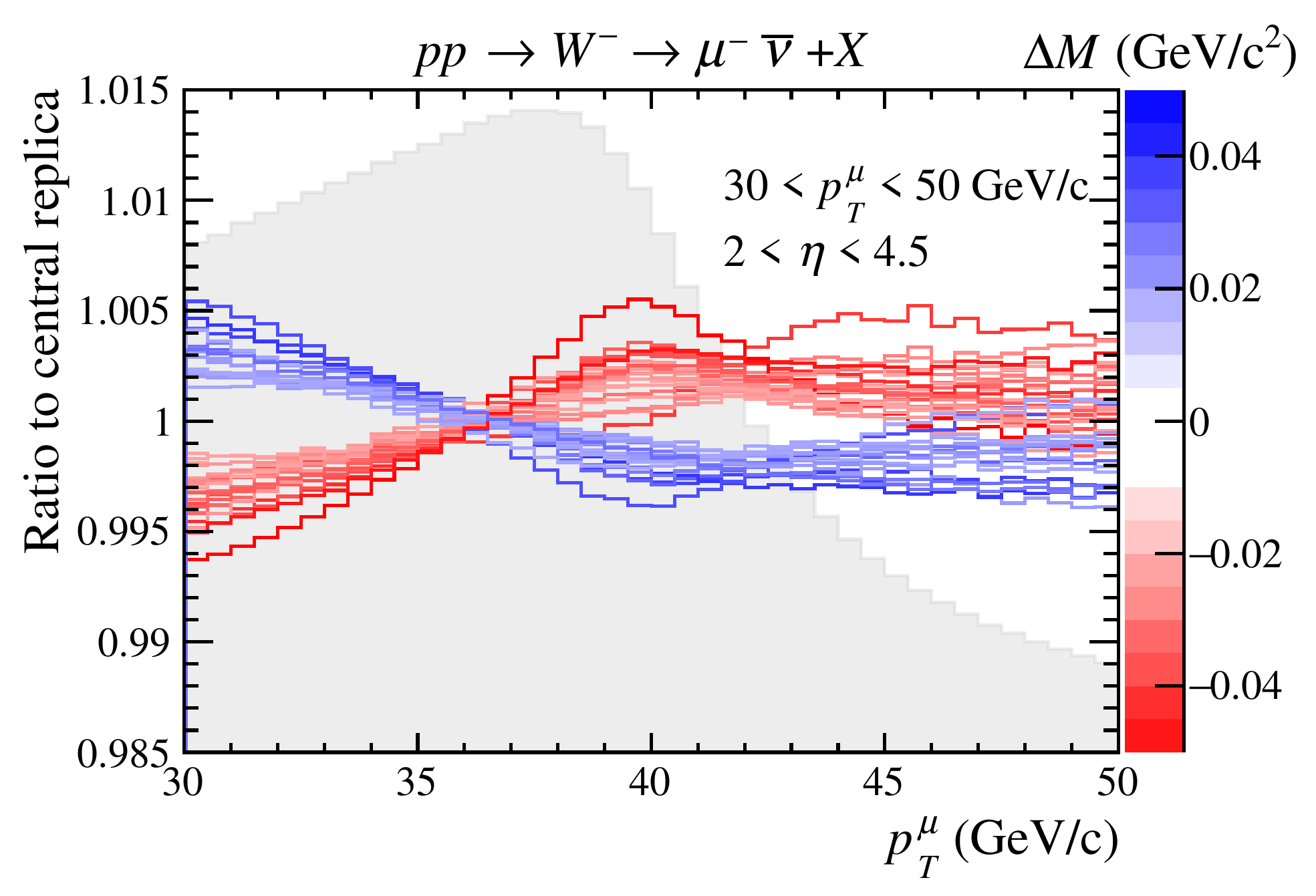}\includegraphics[width=0.46\textwidth]{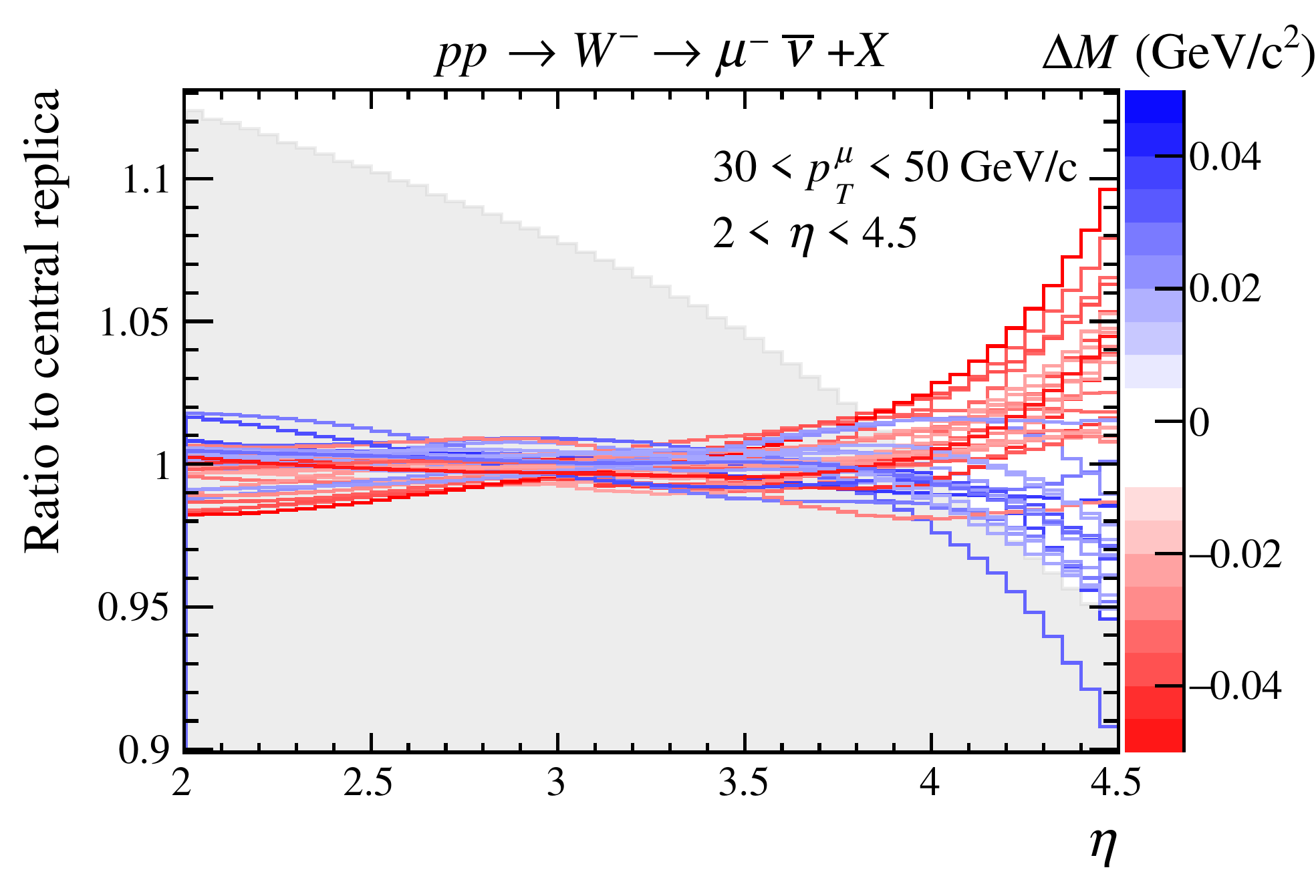}}
\caption{The variations in the shapes of the $p_T^\mu$ and $\eta$ distributions predicted with a subset of NNPDF3.1 replicas. Each line is marked with a colour indicating the shift of the $M_W$ value determined from a fit to the $p_T^\mu$ distribution of a single toy dataset.}\label{fig:4}
\end{figure}

Biases in the determination of $M_W$ are strongly correlated with a mismodelling of the $W$ kinematics. These are characterised by the $W$ transverse momentum, rapidity and polarisation.  Ref.~\cite{PDFpaper} studied the sensitivity of $M_W$ to these variables and to all their possible combinations. However, since these quantities are not directly measurable with LHCb, it is interesting to study the muon kinematic distributions. 
Fig.~\ref{fig:4} shows how the muon $p_T^\mu$ and $\eta$ distributions vary with the PDF replicas. An intriguing observation, however, is that the replicas that cause the largest bias on $M_W$ change not only the shape of the $p_T^\mu$ distribution but also that of the $\eta$ distribution. This is a measurable change, at the level of several percent, which could be exploited to constrain the PDF uncertainty.

\section{The proposed method}
\begin{figure}
\centerline{\includegraphics[width=0.46\textwidth]{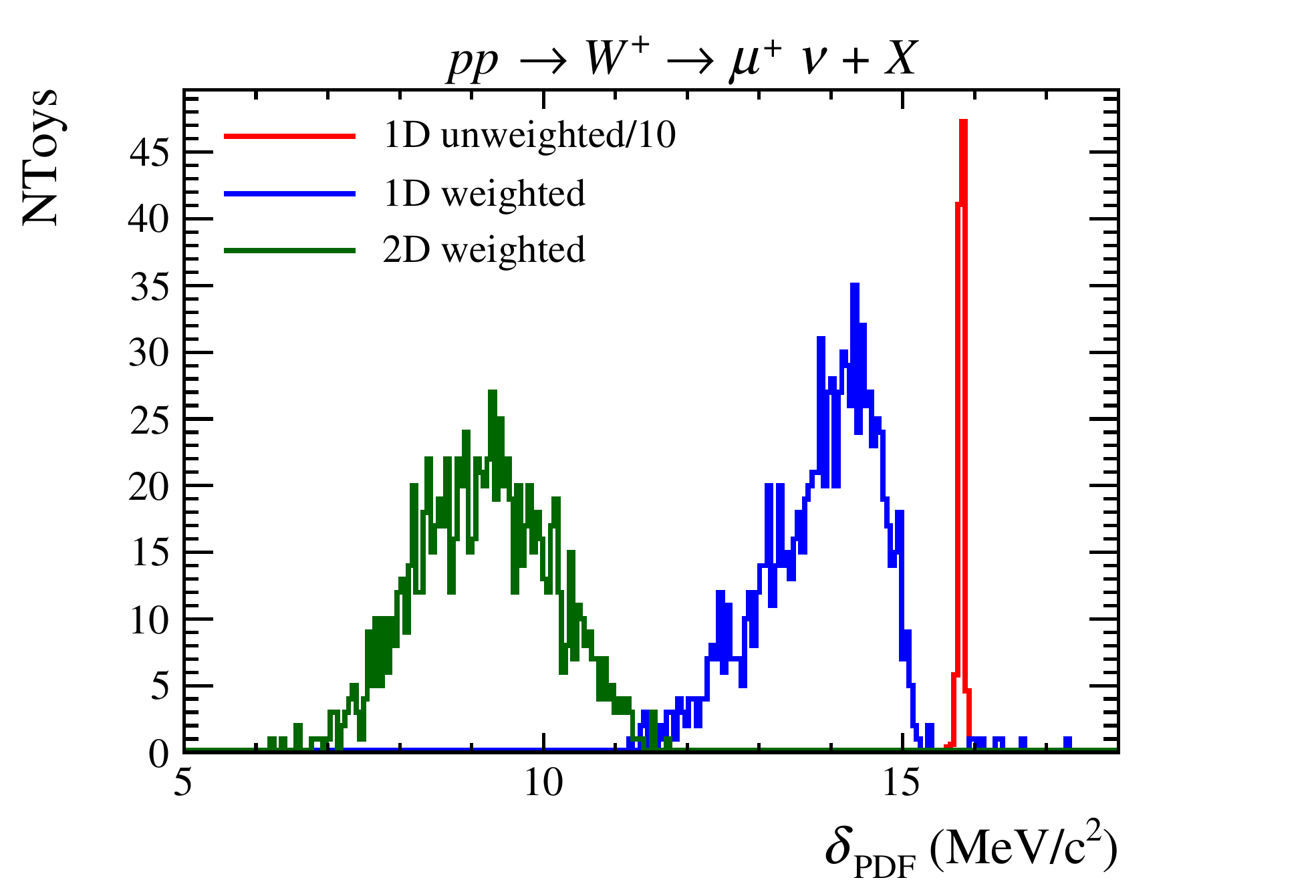}\includegraphics[width=0.46\textwidth]{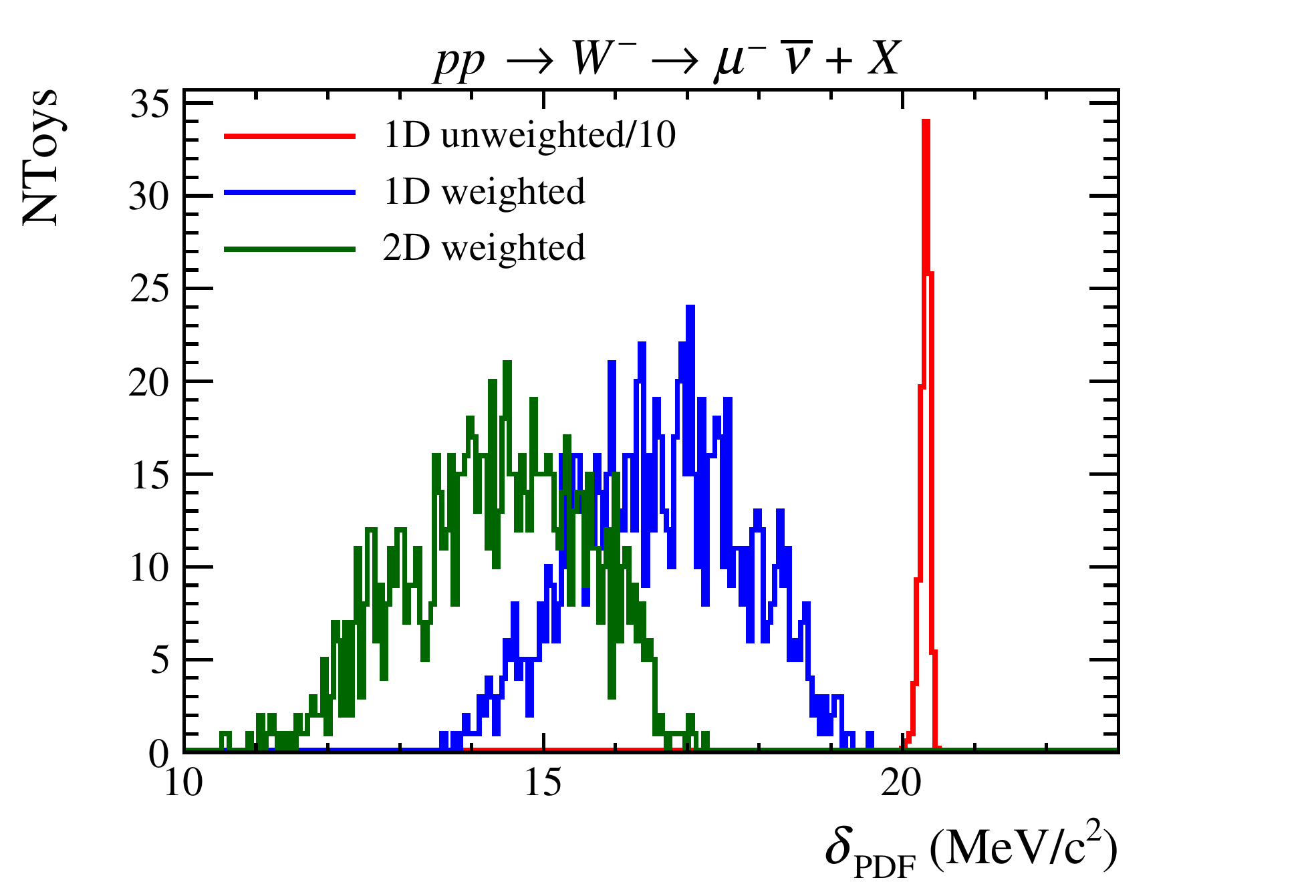}}
\caption{The distribution of the PDF uncertainty evaluated for 1000 toy datasets using three different methods: $p_T^\mu$ fit without weighting, $p_T^\mu$ fit with weighting, ($p_T^\mu , \eta$) fit with weighting. The one-dimensional unweighted distribution is arbitrarily scaled down by a factor of ten.}\label{fig:5}
\end{figure}

Section 3 suggested that a fit to the two-dimensional $p_T^\mu$ versus $\eta$ distribution has potential to further constrain the PDF uncertainty. The traditional one-dimensional fit and the new approach are now compared with the inclusion of PDF replica weights. Each replica is assigned a weight according to the best-fit $\chi^2$ for a fit with $n$ degrees of freedom: $P(\chi^2_{min})\propto \chi_{min}^{2^{(n-1)}}e^{-\chi^2_{min}}$. 
This has the effect of disregarding replicas that are incompatible with the data. The weights are dependent on the toy data, so it is important to consider the results with multiple toy datasets. For a single toy dataset the PDF uncertainty is defined as the RMS of the $M_W$ values for the 1000 replicas. Fig.~\ref{fig:5} shows the distribution of the PDF uncertainty for 1000 toy datasets: the one-dimensional fit with and without weights, and the two-dimensional fit with weights are compared. The largest reduction is observed with the two-dimensional weighted fit, for both the $W$ charges. 

\subsection{Simultaneous fit}
\begin{figure}
\centerline{
\includegraphics[width=0.46\textwidth]{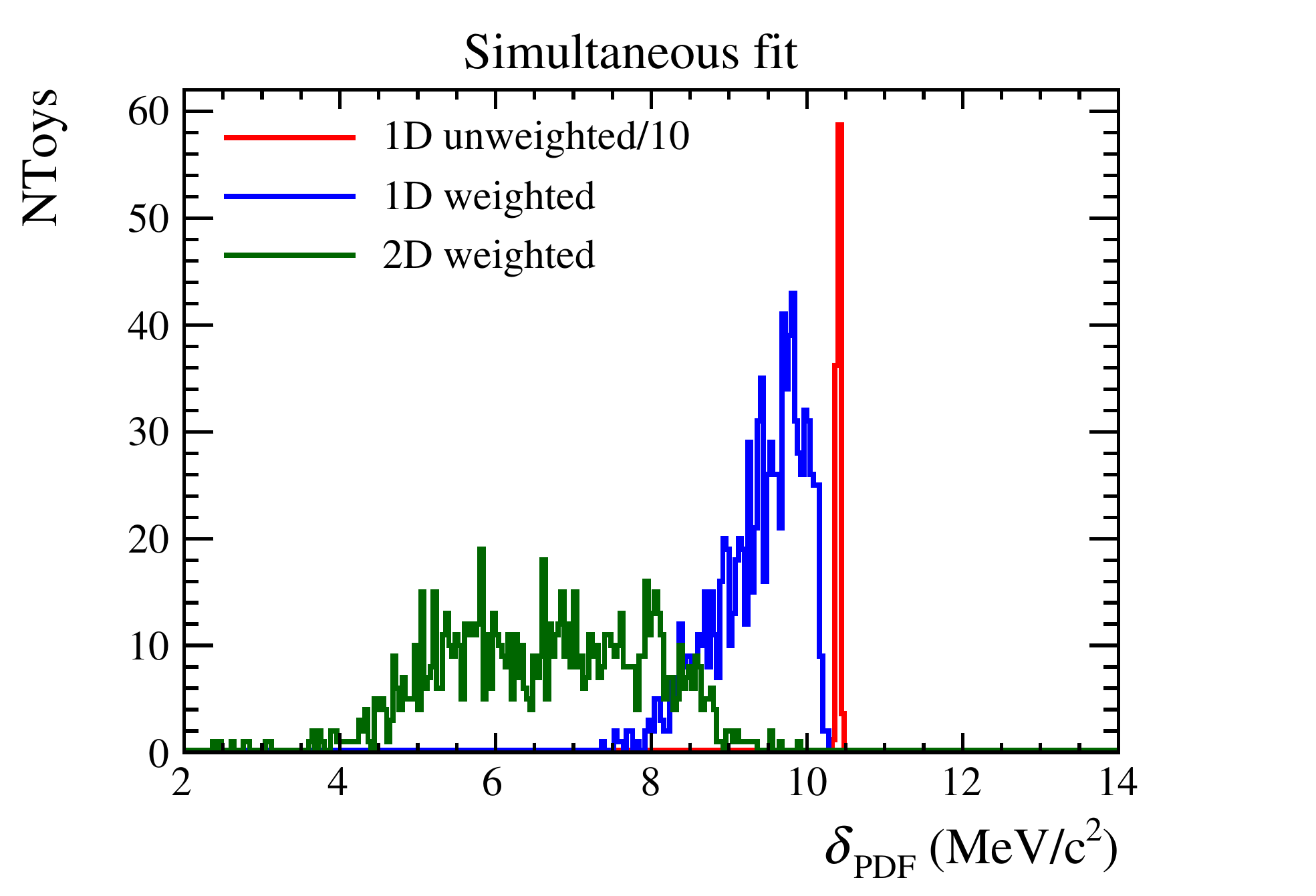}\includegraphics[width=0.46\textwidth]{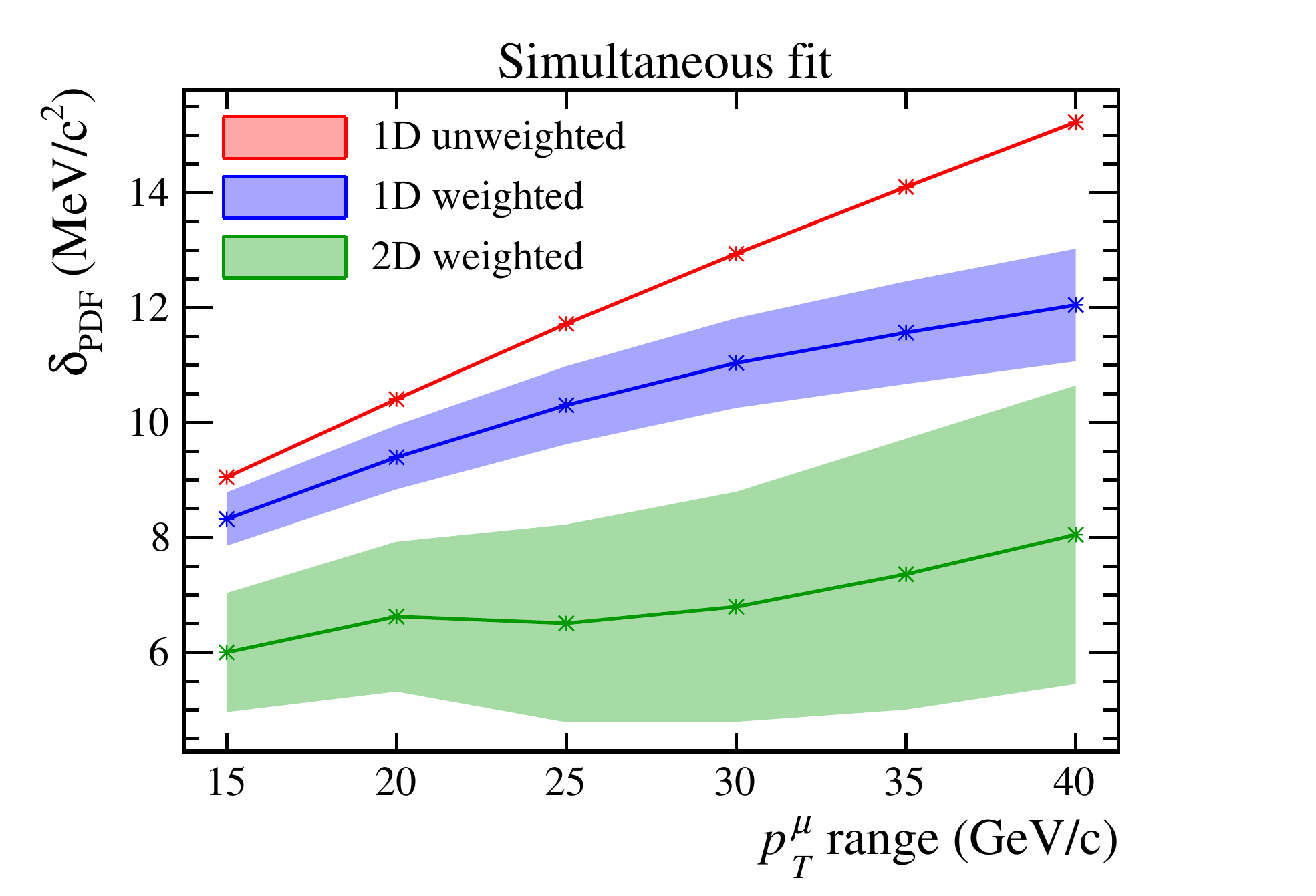}}
\caption{Left: The distribution of the PDF uncertainty evaluated for 1000 toy datasets using a simultaneous fit. Right: PDF uncertainty as a function of the $p_T^\mu$ range used in the simultaneous fit. The bands report the mean and the RMS of the distribution of the PDF uncertainty evaluated for 1000 toy datasets.}\label{fig:6}
\end{figure}

Given the results shown in the previous section it is now interesting to consider the combination of the two charges by performing a simultaneous fit. The left hand side plot in Fig.~\ref{fig:6} shows the PDF uncertainty distribution evaluated for 1000 toy datasets, in which the normalisation of both the $W^+$ and $W^-$ sample is scaled by the same parameter to take into account the integrated charge asymmetry constraint on the PDFs. Compared to the traditional one-dimensional fit, the addition of the weighting typically improves the PDF uncertainty by around 10\%. The two-dimensional fit with weighting is, however, typically around a factor of two better.

\subsection{Dependence on the detector acceptance}
The study was so far restricted to events with 30 $< p_T^\mu < $ 50\,GeV/c and 2 $<\eta<$ 4.5, but it is important to consider how the results depend on this choice. The right hand side plot of Fig.~\ref{fig:6} shows how the PDF uncertainties depend on the width of the $p_T^\mu$ interval, which is symmetric around $M_W/2$.  Each band is centered on the mean of the distribution of the PDF uncertainty evaluated for 1000 toy datasets and its width is defined as the RMS of the same distribution. These results enforce the power of the two-dimensional fit with weighting. A similar study is performed by varying the upper and lower $\eta$ limits~\cite{PDFpaper}. 

\section{Conclusions}
%\vspace{-3mm}
A characterisation of the PDF uncertainty in a future measurement of $M_W$ with LHCb is performed in~\cite{PDFpaper}. A $p^\mu_T$ versus $\eta$ fit with PDF replica weighting can reduce the PDF uncertainty by roughly a factor of two with respect to a simple $p_T^\mu$ fit (the yields are assuming the LHCb Run 2 dataset). Improvements are observed in both the standalone and the simultaneous $W^+$ and $W^-$ fit. 
The new fit approach should be tested on real data: understanding of the muon efficiencies dependence on the $\eta$ spectrum shape will be, therefore, crucial. The accurate modelling of the $p_T^W$ spectrum, the muon momentum scale determination and the control of the background are only some of the challenges that the future measurement will face.

\end{document}